\newcommand{\be}{\begin{equation}}
\newcommand{\ee}{\end{equation}}
\newcommand{\ber}{\begin{eqnarray}}
\newcommand{\eer}{\end{eqnarray}}
\newcommand{\gsim}{\raisebox{-0.7ex}{$\stackrel{\textstyle >}{\sim}$ }}

\documentstyle[preprint,epsfig,epsf,aps,floats]{revtex}
\begin{document}
\tighten
\preprint{\vbox{
\hbox{INT-PUB 02-30}}}
\bigskip
\title{Mixed Kaon Condensation in CFL Matter}
\author {Andrei Kryjevski \footnote{abk4@u.washington.edu}  
and Travis Norsen \footnote{norsen@phys.washington.edu}}
\address{Dept. of Physics and Institute for Nuclear Theory, 
University of Washington, Seattle, WA 98195}
\date{\today}
\maketitle

\begin{abstract}
Previous work on high-density QCD suggests that the ground 
state configuration in the presence of a non-zero neutrino chemical 
potential consists of CFL matter plus a homogeneous condensate
of neutral and positive kaons.  We consider here the stability of 
this homogeneous configuration toward the production of a 
charge-separated mixed phase, and find that the system is indeed stable
against the production of such a heterogenous configuration.
We identify the critical value of the fine structure constant
which would allow for phase separation.  We also derive dispersion
relations for the low-lying excitations above the kaon-condensed
ground state and discuss their possible phenomenological 
implications.
\end{abstract}


\section {Introduction}

It is now understood that at asymptotically large densities, the ground
state of 3-flavor, massless QCD is the Color-Flavor Locked (CFL) 
configuration  \cite{arw,rssv,son}.  In this phase, quarks of all three 
flavors near the (shared) Fermi surface undergo BCS-like pairing due to 
the attractive one gluon exchange potential.  The resulting condensate
breaks the $U(1)_B$ symmetry associated with conserved baryon number, 
and also breaks the original $SU(3)_{color} \times SU(3)_L \times 
SU(3)_R$ symmetry down to the diagonal subgroup, $SU(3)_{C+L+R}$, 
causing the gauge bosons of the original $SU(3)_{color}$ group to 
become massive.  However, a linear combination of the eighth gluon 
and the gauge boson of the $U(1)_{Q}$ symmetry (the photon) remains 
massless;  this unbroken gauge symmetry will be referred to as 
$U(1)_{\tilde{Q}}$, denoting physically a ``rotated'' electromagnetism.  

It was originally suggested that due to the large value of the gap 
the number densities of the three quark flavors would remain equal 
even when perturbations (non-zero quark masses and charge chemical 
potential) were included, since the energy cost of breaking pairs 
exceeded the energy gained by allowing the Fermi spheres for different 
flavors to relax past one another \cite{RW,paulo}.  It was soon realized, 
however, that there exists an energetically cheaper mechanism by which 
the CFL phase could relax under the influence of stresses: meson 
condensation  \cite{sonsteph,bs}.  For example, turning on 
a realistic mass for the strange quark encourages the system
to reduce the strangeness density relative to the density of up- and
down-type quarks. Kaon condensation allows the strangeness 
density to be decreased without the costly breaking of pairs in the
CFL background.  

The phase diagram for CFL + meson condensates was explored in detail
in Ref. \cite{kaplanreddy}.  It was found there that the symmetric
CFL phase was favored only when the chemical potential for electric
charge satisfied $|\mu_Q| \lesssim 10 MeV$ and the quark chemical
potential was many orders of magnitude greater than that expected
in neutron star cores.  At realistic densities, the ground state
at zero lepton chemical potential ($\mu_{\nu}$) involves a condensate 
of $K^0$ while at very high $\mu_{\nu}$  it is the positively charged 
kaon $K^+$ which condenses.  This ensures overall electric charge
neutrality in the presence of a dense background of electrons.

At intermediate $\mu_{\nu}$ it was found that a homogeneous mixed 
phase consisting of both $K^0$ and $K^+$ condensates was most
effective in lowering the free energy.  This mixture sits at a 
free energy saddle point in the hadronic sector configuration space, 
but this saddle point becomes a global minimum when one includes the
free energy contribution of leptons and imposes local electric
charge neutrality.  

One may expect, however, that this homogeneous mixed phase could
be unstable toward the formation of a heterogenous mixed phase in
which the meson condensate rotates toward the $K^0$ condensed phase
in certain regions of space, and toward the $K^+$ phase in others.
This would violate the original assumption of {\it local} electric
charge neutrality, since the $K^+$ regions would become
positively charged and the $K^0$ regions would become negatively
charged. This is allowed so long as the volume fractions
of the two phases are such as to conserve charge globally.  The
question of stability is then the question of whether the
cost associated with surface and Coulomb energies can be made 
less than the energy gained by allowing the meson fields to slide
down from the saddle point to regions of lower potential.

This is the issue addressed in this paper, which is organized 
as follows. After a quick review of the chiral Lagrangian formalism 
used to describe mesonic excitations above the symmetric CFL ground state, 
we argue that the standard approach to studying possible mixed phases 
(involving some version of the thin wall approximation) will not be 
appropriate in this case.  We then present the results of a 
small-fluctuation stability analysis of the homogeneous $K^0/K^+$ 
mixed phase to derive the dispersion relations for the
low-lying excitations and discuss their implications to neutron 
star cooling and other possible phenomenological issues.

\section{Meson Condensation in CFL Matter}

Because of the identity between the symmetry breaking pattern in the
vacuum structure of QCD and the symmetry breaking pattern in the CFL
phase at high density, it is appropriate to use the same chiral 
Lagrangian description one uses to parametrize small excitations
about the QCD vacuum to parametrize small fluctuations about the
$SU(3)$-symmetric CFL phase. 

These low energy excitations are parametrized by
$\Sigma = e^{2i\pi/f_{\pi}+\eta^{'}/f_{A}}$ and $B=e^{\beta/f_{B}}$ (with 
$\pi = \pi^a T^a$) where $\pi_a$'s are the pseudo-scalar octet
of Goldstone bosons which arise from the breaking of chiral symmetry 
and $T_a$'s are Gell-Mann matrices. $\eta^{'}$ and $\beta$ are Goldstone 
bosons due to breaking of the $U(1)_A$ (assumed to be a valid symmetry of 
the theory at high density) and $U(1)_{B}$ symmetries, respectively. 
Under the original symmetry of $SU(3)_{L}\times SU(3)_{R}\times U(1)_{B}
\times U(1)_{A}$ the $\Sigma$ field transforms as $(3,\bar3)_{0,-4}$.

The quark mass matrix $M={\mathrm diag}(m_u,m_d,m_s)$ induces mass 
differences between the members of the octet and transforms as 
$(3,\bar3)_{0,2}$.  Due to $U(1)_{A}$ charges only even powers of 
$M$ are present in the effective Lagrangian.  To the leading order, 
only the combination $\tilde{M} = |M| M^{-1} = 
{\mathrm diag}(m_d m_s, m_u m_s, m_u m_d)$ that transforms as 
$(\bar3,3)_{0,4}$ contributes to explicit mass terms via the operator
${\mathrm Tr} \Sigma \tilde{M}$.  The operator $X= -M^{\dag}M / 2\mu$ 
acts as a chemical potential for flavor symmetries \cite{bs}.  Here 
$\mu$ represents the quark chemical potential.  The leading order 
chiral Lagrangian for the octet is then:
\be
{\mathcal L} = {f^2_{\pi}} \left[{1\over{4}} {\mathrm Tr} \left( D_0 
\Sigma D_0 \Sigma^{\dag} - v^2 {\mathrm Tr} \vec{\nabla} \Sigma
\cdot \vec{\nabla}\Sigma \right) + {a\over{2}} {\mathrm Tr} \tilde{M}
(\Sigma + \Sigma^{\dag}) + {b\over{2}} {\mathrm Tr} Q \Sigma Q 
\Sigma^{\dag} \right]
\label{lagrangian}
\ee
where $v^2=1/3$ is the (squared) in-medium meson velocity.    
The covariant derivative 
is given by
\be
D_0 \Sigma = \partial_0 \Sigma - i [\tilde{\mu},\Sigma]
\ee
where $\tilde{\mu} = \mu_Q Q + X$ and $\mu_Q$ is the chemical
potential for electric charge and $Q={\mathrm diag}(2/3,-1/3,-1/3)$.  
The coefficient $a$ has been computed in Ref. \cite{sonsteph,beane} 
and is given by $a = \frac{3 \Delta^2}{\pi^2 f^2_{\pi}}$.  The 
coefficient $b$, which accounts for charge-dependent differences 
in the meson masses, has been previously estimated by dimensional 
analysis to be of order $b \sim \frac{\alpha \Delta^2}{4\pi}$ where 
$\alpha$ is the fine structure constant corresponding to the unbroken 
$U(1)_{\tilde{Q}}$ in the CFL phase.  \cite{kaplanreddy,manuel}
The parameter $\Delta$ denotes the value of the gap, which we will 
always approximate by $\Delta = 100 MeV$.

The free energy density arising from Eq.(\ref{lagrangian}) relative
to the free energy energy density of the symmetric CFL phase is:
\be
\Omega = \frac{f^2_{\pi}}{4} \left[ {\mathrm Tr} [\tilde{\mu},\Sigma]
[\tilde{\mu},\Sigma^{\dag}] + v^2 {\mathrm Tr} \vec{\nabla}\Sigma
\cdot \vec{\nabla}\Sigma - 2 a {\mathrm Tr} \tilde{M}(\Sigma +
\Sigma^{\dag} - 2) - b {\mathrm Tr} [Q,\Sigma][Q,\Sigma^{\dag}] 
\right]
\label{freeenergy}
\ee
where $\Sigma$ is an $SU(3)$ matrix characterizing the 
direction of the condensate in internal space.  As shown in Ref.
\cite{kaplanreddy} all stationary points of the free energy for spatially
homogeneous configurations will satisfy the matrix equation
\be
0 = \left[ \tilde{\mu}\Sigma^{\dag} \tilde{\mu} \Sigma - a \tilde{M}
\Sigma - b Q \Sigma^{\dag} Q \Sigma \right] - h.c. 
\label{stationary}
\ee
This equation has two solutions corresponding to minima of Eq. 
(\ref{freeenergy}).  The first such minimum corresponds to a 
$K^0$ condensate and is parametrized by
\be
\Sigma = {\mathrm exp} \left[ i\, \theta_{K^0} \left( \begin{array}{ccc}
0 & 0 & 0 \\ 
0 & 0 & 1 \\
0 & 1 & 0 \end{array}
\right) \right]
\ee
where $\theta_{K^0}$ is given by
\be
{\mathrm cos}\,\theta_{K^0} = \frac{M^2_{K^0}}{\tilde{\mu}^2_{K^0}}.
\ee
Where ${M^2_{K^0}}={\mathrm a}(m_d+m_s)m_u$ and $\tilde{\mu}^2_{K^0}=({\mathrm} m_s^2-m_d^2)/2\mu$ \cite{kaplanreddy}.
(The minimum exists whenever this angle is real.)  Similarly, 
there exists a local minimum 
of the free energy corresponding to a pure $K^+$ condensate:
\be
\Sigma = {\mathrm exp} \left[ i\, \theta_{K^+} \left( \begin{array}{ccc}
0 & 0 & 1 \\
0 & 0 & 0 \\
1 & 0 & 0 \end{array}
\right) \right]
\ee
where $\theta_{K^+}$ is given by
\be
{\mathrm cos}\,\theta_{K^+} = \frac{M^2_{K^+} - b}{\tilde{\mu}^2_{K^+}-b}
\ee
where ${M^2_{K^+}}={\mathrm a}(m_u+m_s)m_d$ and $\tilde{\mu}^2_{K^+}=
\mu_{Q}+({\mathrm} m_s^2-m_u^2)/2\mu$
and again the solution only exists when $\theta_{K^+}$ is real.

Between the two minima just mentioned, there exists a free energy
saddle point configuration involving a mixed condensate of $K^0$
and $K^+$.  Parameterizing the condensate by
\be
\Sigma = {\mathrm exp} \left[ i\, \theta_{SP} \left( \begin{array}{ccc}
0 & 0 & {\mathrm sin} \phi \\
0 & 0 & {\mathrm cos} \phi \\
{\mathrm sin} \phi & {\mathrm cos} \phi & 0 \end{array}
\right) \right]
\label{ansatz}
\ee
one finds a solution to Eq. (\ref{stationary}) with angles satisfying:
\be
{\mathrm cos}\,\theta_{SP} = \frac{M^2_{K^0}}{\tilde{\mu}^2_{K^0}} +
\frac{\tilde{\mu}_{K^+}(\tilde{\mu}_{K^+}-\tilde{\mu}_{K^0})}
   {b\tilde{\mu}_{K^0}}
\left[ \left( \frac{M^2_{K^+}-\tilde{\mu}^2_{K^+}}{\tilde{\mu}_{K^+}}\right)
 - \left( \frac{M^2_{K^0}-\tilde{\mu}^2_{K^0}}{\tilde{\mu}_{K^0}}\right)
\right]
\label{theta}
\ee
and
\be
{\mathrm sin}^2 \phi = \frac{ \left( \frac{M^2_{K^0}}{\tilde{\mu}^2_{K^0}} -
 {\mathrm cos}\,\theta_{SP} \right)}{ \left( 1 - \frac{\tilde{\mu}_{K^+}}
   {\tilde{\mu}_{K^0}} \right) \left( 1 - {\mathrm cos}\,\theta_{SP}
\right)}
\label{phi}
\ee
where as usual the solution is only a stationary point of the free energy
if the angles $\theta_{SP}$ and $\phi$ are real.

It was found in Ref. \cite{kaplanreddy} that this saddle point 
configuration becomes the true minimum of the free energy when one
includes leptonic contributions to the free energy and imposes 
local electric charge neutrality ({\it i.e.}, homogeneity) on the
system.  In practice, this means that for a given quark chemical 
potential $\mu$ and lepton chemical potential $\mu_{\nu}$, the 
chemical potential for electric charge is determined by the 
constraint that the electron number density equal the number density
of $K^+$ in the meson condensate:
\be
\frac{(\mu_{\nu}-\mu_Q)^3}{3 \pi^2} = -\frac{\partial \Omega_{meson}}
{\partial \mu_Q}
\ee
where $\Omega_{meson}$ is given by Eq. (\ref{freeenergy}).  

\section{Heterogenous Meson Condensation}

It is well-known that the imposition of local electric charge 
neutrality is often unwarranted, and may hide a lower-energy configuration
in which overall charge neutrality results from the cancellation of
charge among regions of non-zero local charge density  
\cite{glendenning}.  If this occurs, the resulting crystalline mixed 
phase may dramatically affect both the equation of state and various 
transport properties of the neutron star \cite{hps}.

\begin{figure}[t]
\centering{
\epsfig{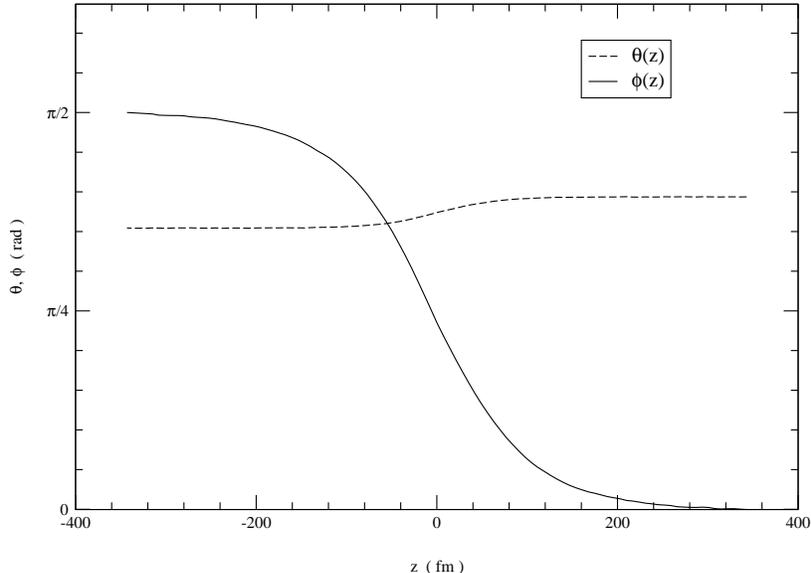}
}
\caption{Minimal-surface-tension boundary profile between regions of
pure $K^0$ condensed matter and pure $K^+$ condensed matter at a quark
chemical potential $\mu = 400MeV$.  The bulk phases on either side
of the boundary are at equal pressures.  The effects of static Coulomb
screening are here ignored.  The surface tension found by integrating
the excess free energy density across the boundary is
$\sigma = 0.082 MeV/fm^2$.}
\label{fig1}
\end{figure}

It is natural to wonder whether the assumption of local charge
neutrality mentioned in the previous section was overly restrictive.
We begin our analysis of this question by assuming that an inhomogeneous
mixed phase of $K^0$ and $K^+$ condensed regions is indeed favored,
and studying the nature of the boundary layer between two adjacent
regions. For now we ignore electrostatic energy contributions
and Coulomb screening effects.  Following the approach of Ref. 
\cite{surface} we consider a planar boundary and fix the condensates 
to point in the $K^0$ and $K^+$ directions at $\pm \infty$ respectively.  
We then implement a relaxation procedure which allows the fields to 
interpolate between the two fixed boundaries in a way that minimizes 
the total free energy of the surface.  Pressure equilibrium across the 
boundary imposes a constraint which fixes the value of $\mu_Q$.  The 
resulting  boundary profile, parametrized by the angles $\theta$ and 
$\phi$ of Eq. (\ref{ansatz}), for a quark chemical potential 
$\mu = 400 MeV$ is shown in Fig. \ref{fig1}.  

The notable feature is the large thickness of the energetically favored
boundary.  One can understand it by comparing the two terms contributing 
to the excess free energy of the boundary:  the gradient terms associated 
with spatial changes in the field configurations, and the ``bulk'' terms 
which tally the cost of being away from a free energy minimum while one
traverses the boundary.  Since the height of the free energy barrier 
between the two minima is very small (of order $10^3 MeV^4$), a very 
thick boundary is required in order to make the two contributions 
comparable, and hence to minimize the overall surface tension.  In the 
example shown, the surface tension turns out to be the quite small 
value $\sigma = 0.082 MeV/fm^2$.

However, a boundary layer between $K^0$ and $K^+$ regions with thickness
of several hundred $fm$ is not allowed physically, since the Debye screening
length of all charged species is much smaller than this length scale.
Any structure of this enormous size would be warped out of existence 
as charge flowed in response to the large electric fields. We can 
estimate the role of screening by considering the energetics of a 
single charged droplet of the charged, lower-energy-density phase immersed
in a background of the original phase.  Ignoring for the moment the
Coulomb contribution to the energy, we find
\be
E(R) \sim \frac{4}{3}\pi \Delta \epsilon R^3 + 4 \pi \sigma R^2
\ee
where $\Delta  \epsilon \le 0$ is the bulk (free) energy density 
difference of the two phases.  Any droplet larger than $R_{crit} \sim 
3 \sigma / |\Delta  \epsilon|$ will be energetically favored compared to the
homogeneous phase, and the volume and surface terms in the energy are
of comparable size for droplets of radius near $R_{crit}$.  In order
for the droplet to be stable against charged particle flow
(Debye screening), the Coulomb contribution to $E(R)$ must be a small
correction to the terms already written.  Estimating the Coulomb
term by $E_C(R) \sim \alpha \rho_Q^2 R^5$ (where $\rho_Q$ is the
electric charge density of the droplet) then gives
\be
\alpha \ll \frac{|\Delta \epsilon| ^3}{\sigma^2 \rho_Q^2}
\ee
as the condition for the existence of a charge-separated mixed
phase.  In the present case, the right hand side 
is of order $10^{-8}$ and a stable heterogenous
mixed phase (in which each phase sits at one of the local minima of
the free energy density) is clearly ruled out for the physical
value $\alpha \sim 1/137$.

If a heterogenous mixed phase is formed, then, it will have to involve
only smaller departures from the homogeneous saddle point configuration.
The condensate may ``slide down'' toward the $K^0$ configuration in
one region of space, and similarly ``slide down'' toward the $K^+$
configuration in an adjacent region, but it cannot slide all the way
down to the free energy local minimum.  And without any definite 
guess as to the bulk field configurations in each region, it is 
difficult to study the energetic favorability of possible mixed phases 
in the standard way -- that is, by taking an ansatz for the mixed phase 
configuration, treating the boundary as ``thin'', varying with
respect to appropriate parameters (droplet size, {\it etc}.), and 
comparing the total free energy to the free energy of the homogeneous
phase.

We therefore take a somewhat different approach and study the stability
of small perturbations to the homogeneous saddle point configuration.
Any unstable modes will signal the instability of the homogeneous phase 
toward the production of some (largely unspecified) heterogenous mixed 
phase.

We present the full results of our analysis of the excitation spectrum
in the next section; here we will simply quote those results which
are relevant to the question of the stability of the homogeneous mixed
phase.  As we show in detail in Sec. \ref{sec4} the kaonic excitations about
the homogeneous, charge neutral configuration consist of two massive
kaons, one massless Goldstone boson arising from the breaking of 
$U(1)_Y$, and a massive in-medium longitudinal photon (which acquired 
a mass through the Higgs mechanism by eating one of the kaonic degrees
of freedom).  Plane-wave oscillations of this longitudinal photon degree
of freedom represent longitudinal charge density waves -- just the sort
of excitations which will be unstable if the homogeneous phase can lower
its energy by forming a charge-separated mixed phase.

\begin{figure}[tb]
\centering{
\epsfig{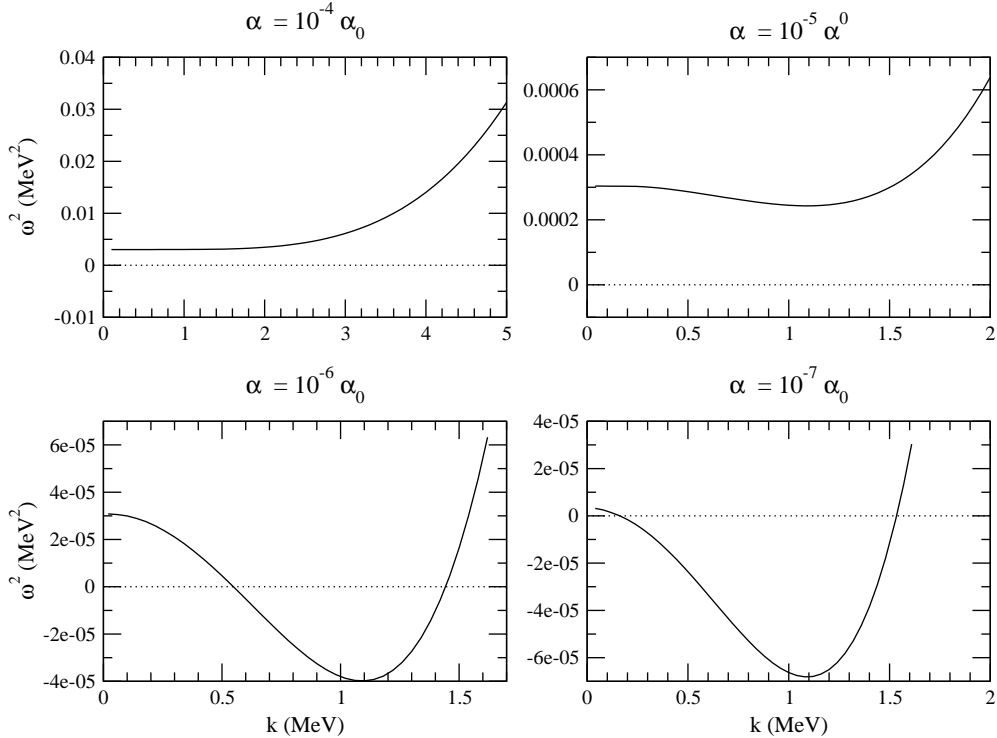}
}
\caption{Dispersion curves for the longitudinal photon mode for several
values of $\alpha$.  ($\alpha_0$ represents the value $1/137$.)  The
quark chemical potential is set to $\mu = 400 MeV$ while $\mu_{\nu} =
100 MeV$.  For $\alpha \le 10^{-6} \alpha_0$ the energy of the mode 
becomes pure imaginary (for a certain range of wavelengths),
corresponding to an instability.  Debye screening by electrons is
included by giving the longitudinal photon an additional mass squared
term equal to $k_D^2 = q^2 \mu_e^2 / (2 \pi^2)$.  Note also that the 
Higgs contribution to the photon mass is proportional to the electric 
charge $q$; hence the overall reduction in scale as $\alpha$ is reduced.}
\label{fig2half}
\end{figure}

In Fig. \ref{fig2half} we show the dispersion relation for this 
mode as the electric charge $\alpha$ is varied.  The longitudinal 
charge density wave is clearly a stable, massive mode until $\alpha$ 
is reduced many orders of magnitude below the physical value.  Note 
that the critical value of $\alpha$ is not too far from the naive 
estimate made above based on consideration of macroscopic droplet 
properties.  Also, the static plasma wave is most unstable for wave
vectors
$|{\bf k}| \sim 1 MeV$ which corresponds to charged structures with a size 
of $D \sim 1/|{\bf k}| \sim 200 fm$.  This is again comparable to the size 
expected from the minimal-surface-tension boundary constructed above 
and shown in Fig. \ref{fig1} -- and is safely smaller than the
charged particle Debye screening lengths for such tiny $\alpha$.
We conclude that in real high density matter where $\alpha \sim 1/137$
the homogeneous $K^+/K^0$ mixed phase is indeed stable 
against the production of a heterogenous, charge-separated 
configuration.  The energetic cost of Coulomb and surface/gradient
energy wins out over the energy savings associated with lowering
the bulk energy densities.

\section{Excitation Spectrum}
\label{sec4}

Having established that the homogeneous $K^+/K^0$ mixed is indeed the
true ground state of matter at high density and nonzero neutrino
chemical potential, we are now interested in finding the normal modes 
and dispersion relations for excitations above the ground state.
The homogeneous background condensate is given by
\be
\Sigma_0 = {\mathrm exp} \left[ i\, \theta_{SP} \left( \begin{array}{ccc}
0 & 0 & {\mathrm sin} \phi \\
0 & 0 & {\mathrm cos} \phi \\
{\mathrm sin} \phi & {\mathrm cos} \phi & 0 \end{array}
\right) \right]
\ee
where the angles are given by Equations (\ref{theta}) and (\ref{phi}).  
In order to parameterize small fluctuations about this state, we
write the condensate as
\be
\Sigma = \xi_0\, \hat{\Sigma}\, \xi_0
\label{ansatz2}
\ee
where $\xi_0 = \Sigma_0^{1/2}$ and
\be
\hat{\Sigma} = {\mathrm exp}(2i\hat{\pi}/f_{\pi}).
\ee
Here $\hat{\pi} = \hat{\pi}_a T_a$ characterizes small rotations away
from the homogeneous saddle point configuration given by $\Sigma_0$.
\footnote{This helpful parameterization was suggested by David Kaplan.}

In order to account for the Coulomb energy 
associated with possible charge separation, we must include the coupling
of our meson fields to the gauge field associated with the unbroken
$U(1)_{\tilde{Q}}$.  This is accomplished by promoting the
electric charge chemical potential to the role of a space- and
time-dependent field.  (For the following analysis,
we will work in the Coulomb gauge $\vec{\nabla} \cdot {\bf{A}}=0$
in which, to quadratic order in the Lagrangian, the transverse 
photon modes decouple from the hadronic excitations.  Hence for our
purposes we only need to keep track of the electrostatic potential.
The transverse photon modes will acquire a mass via the Higgs 
mechanism in the standard way.)  In particular, since we are concerned 
only with small perturbations about the homogeneous phase, we replace
\be
\mu_Q \rightarrow \mu_Q + \delta\! A_0({\bf{x}},t)
\label{emfield}
\ee
where $\mu_Q$ is the constant value found by requiring the homogeneous
phase to be electrically neutral, and $\delta\! A_0$ characterizes local
fluctuations in the electrostatic potential.  Note that from now on,
we ignore the distinction between $Q$ and $\tilde{Q}$ charges.  This is
expected to be physically accurate for accessible densities, where the
mixing angle between the new and old electromagnetism is small 
\cite{interface}.

Substituting Equations (\ref{ansatz2}) and (\ref{emfield}) into our 
Lagrangian, and keeping only terms bilinear in fields we arrive at:
\ber
{\mathcal L} &=& {\mathrm Tr} \left( \partial_t \hat{\pi}
\partial_t \hat{\pi} - v^2 (\vec{\nabla}
 \hat{\pi})^2 \right)
-i {\mathrm Tr} \partial_t \hat{\pi} \left(
   [\mu_R , \hat{\pi}] + [\mu_L , \hat{\pi}] \right) \nonumber \\
&&+f_{\pi} q \delta\! A_0 {\mathrm Tr} \partial_t \hat{\pi}
   (Q_R - Q_L)
- i f_{\pi} q \delta\! A_0 {\mathrm Tr}\left( [Q_R , \mu_L]
   - [Q_L , \mu_R]\right)\hat{\pi} \nonumber \\
&& -\frac{1}{2}f^2_{\pi} q^2 \delta\! A_0 ^2 {\mathrm Tr} (Q_R Q_L - Q^2)
- {\mathrm Tr} [\mu_L , \hat{\pi}][\mu_R , \hat{\pi}] \nonumber \\
&& - a {\mathrm Tr}(M_L + M_R)\hat{\pi}^2
   + b {\mathrm Tr}[Q_L,\hat{\pi}][Q_R,\hat{\pi}] \nonumber \\
&& + \frac{1}{2}(\vec{\nabla}\delta\!A_0)^2 - 
\frac{1}{2}\delta\! A_0 ^2 \Pi^{00}_{elec}(\omega,k)
\label{quadL}
\eer
where we have defined:
\ber
\mu_R &=& \xi_0 \tilde{\mu} \xi_0^{\dag} \nonumber \\
\mu_L &=& \xi_0^{\dag} \tilde{\mu} \xi_0 \\
M_R &=& \xi_0^{\dag} \tilde{M} \xi_0^{\dag} \nonumber \\
M_L &=& \xi_0 \tilde{M} \xi_0 \\
Q_R &=& \xi_0 Q \xi_0^{\dag} \nonumber \\
Q_L &=& \xi_0^{\dag} Q \xi_0 .
\eer
(Note that the definitions of $M_R$ and $M_L$ differ from the other
two cases.)  In the last line of Eq. (\ref{quadL}) the first term
represents the electrostatic field energy, while the second accounts
for the coupling of the background electron bath to the electric 
potential.

The factor $\Pi^{00}_{elec}(\omega,{\bf k})$ is the contribution to the
temporal photon mass (squared) arising from the background
electron gas.  The leading-order (gauge-invariant) contribution
to this quantity can be written down exactly for the case of a
degenerate, relativistic electron gas \cite{manuel2,braaten}.  
It is:
\be
\Pi^{00}_{elec}(\omega,{\bf k}) = k_D^2 \left( \frac{\omega}{2|{\bf k}|}
  {\mathrm ln}\left| \frac{\omega + |{\bf k}|}{\omega - |{\bf k}|} \right| - 
1 \right)
\ee
where $k_D^2 = \frac{q^2 \mu_e^2}{2 \pi^2}$ is the (squared) electron
Debye screening length.  We have omitted a possible imaginary part which 
describes Landau damping.  We expect this damping to become important 
only for $|{\bf k}| \sim \mu_e$.  Note that in the static limit $\omega 
\ll |{\bf k}|$ the electron Debye contribution to the photon mass squared 
becomes simply the constant $k_D^2$.  This was
the approximation used in the previous section when studying the
stability of the homogeneous configuration.  In the present section,
we retain the full expression.  The effect of the electrons for
$|{\bf k}| \lesssim k_D$, it turns out, is merely to increase the effective 
mass of the temporal photon by $1 - 2 MeV$ above the mass it gets via 
the Higgs mechanism.

\begin{figure}[tb]
\centering{
\epsfig{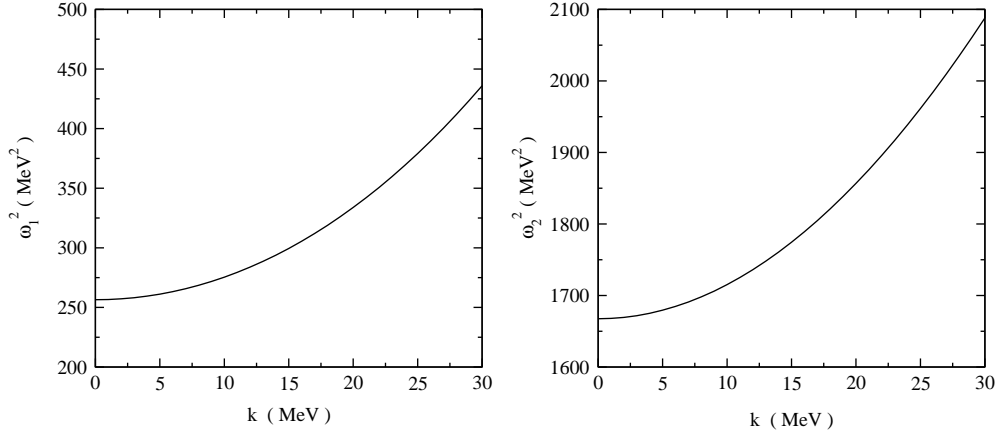}
}
\caption{Dispersion curves for the two charged pion normal modes at
$\mu = 400 MeV$ and $\mu_{\nu} = 100 MeV$.  The
first normal mode corresponds to an almost pure $\pi^-$ state with a
numerically tiny admixture of $\pi^+$; {\it vice versa} for the second 
mode.  The in-medium masses of the two charged pions are 
approximately $15 MeV$ (for the $\pi^-$-like state) and $40 MeV$ (for 
the $\pi^+$-like state).  The large mass splitting between the 
differently charged states arises from the nonzero value of the
charge chemical potential, $\mu_Q$.}
\label{fig2}
\end{figure}

It is now possible to use symmetries to determine the superselection
rules, and make a separate (sparse) ansatz for $\hat{\pi}$ for each
non-interacting group of excitations.  The simplest such group is the
charged pions, $\pi^+$ and $\pi^-$.  We will show the details of this
calculation for pedagogical purposes and omit them in the physically
more interesting case of kaonic excitations, which we treat next.  
Writing the pion fields in terms of their real and imaginary parts we 
have:
\be
\hat{\pi} = \frac{1}{\sqrt{2}}\left( \pi_1 T_1 
                                    +\pi_2 T_2 \right)
\label{chpi}
\ee
where $T_a$ are $SU(3)$ generators with ${\mathrm Tr}(T_a T_b)=
(1/2)\delta_{ab}$.
Inserting Eq. (\ref{chpi}) into the Lagrangian Eq. (\ref{quadL}) and
taking appropriate derivatives yields
three coupled Euler-Lagrange equations for the fields
$\pi^{\pm}_1$, $\pi^{\pm}_2$, and $\delta \! A_0$.  Taking a plane wave 
ansatz for each field
\be
\left( \begin{array}{c} \pi_1 ({\bf x},t) \\ 
                        \pi_2 ({\bf x},t) \\
                        \delta \! A_0 ({\bf x},t) \end{array} \right)
 \rightarrow 
\left( \begin{array}{c} \pi_1 \\ 
                        \pi_2 \\
                        \delta \! A_0 \end{array} \right)
e^{i({\bf k}\cdot{\bf x}-\omega t)}
\ee
then gives a homogeneous linear system in the three field amplitudes.
The solutions of this system represent the normal modes and dispersion
relations.  Since there is no term in the Lagrangian that depends on 
$\partial \delta \! A_0 / \partial t$, the equation of motion for 
$\delta \! A_0$ -- Poisson's equation -- should be interpreted as a 
constraint.  (We are working in Coulomb gauge, where this is standard.)
Hence, the system of three linearized equations of motion 
will have only two normal modes.  The dispersion curves for these two 
modes (which are almost pure $\pi^-$ and $\pi^+$ respectively) 
are shown in Fig. \ref{fig2}.  

\begin{figure}[htb]
\centering{
\epsfig{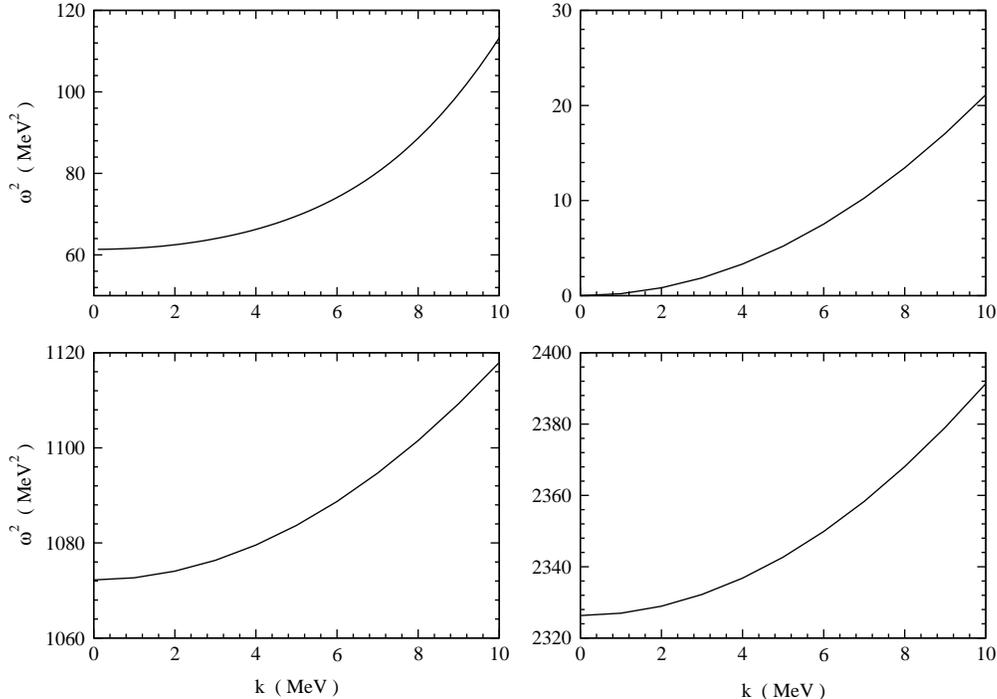}
}
\caption{Dispersion curves for the four kaonic normal modes at
$\mu = 400 MeV$ and $\mu_{\nu} = 100 MeV$.  The
upper-left figure shows the longitudinal photon mode, which we have
also described as a longitudinal plasma oscillation.  With the 
physical value of the fine structure constant, this mode is stable
($\omega^2 > 0$) and the homogeneous $K^0/K^+$ mixed phase is the
true ground state of the system.  There are also two massive kaon
modes, and a single massless mode arising from the breaking of the
$U(1)_Y$ hypercharge symmetry by the $K^0$ portion of the condensate.}
\label{fig3}
\end{figure}

Because we are expanding about a background condensate consisting of
both $K^0$ and $K^+$, the neutral kaonic excitations may mix with
the charged kaons.  In order to find the normal modes and dispersion
relations, we proceed as before, writing
\be
\hat{\pi} = \frac{1}{\sqrt{2}}\left( K^0_1 T_6 
                                    +K^0_2 T_7
                                    +K_1 T_4
                                    +K_2 T_5 \right)
\label{kaonansatz}
\ee
and substituting this ansatz into the quadratic Lagrangian.  Requiring
harmonic dependence on $x$ and $t$ as before gives a coupled system of
five equations of motion for the field amplitudes.  The dispersion
relations for the four normal modes are shown in Fig. \ref{fig3}.

There is one exactly massless mode which can be identified as the
Goldstone boson arising from the fact that the $K^0/K^+$ condensate
breaks the $U(1)_Y$ hypercharge symmetry of the original theory.  
(Weak interactions, however, will give a small mass to this
particle \cite{sonweak}.)  This
Goldstone boson corresponds to $K^0_2$ in our above ansatz, {\it i.e.},
it consists of harmonic fluctuations in the imaginary part of the 
neutral kaon fields (as expected). There are also two massive modes 
(with masses of about  $32 MeV$ and $48 MeV$).  The lighter of these is 
predominantly a $K^0$ with a small admixture of $K^0_1$ and $K_1$, 
while the heavier is mostly a $K^-$ with a small admixture of 
$K^0$.  Finally, the temporal photon mode (consisting, as expected,
of the imaginary part of the charged kaon field plus a sizable electric
field) weighs in with a mass of about $8 MeV$.  As discussed previously,
this mode (which came from the photon eating
the would-be-Goldstone-boson from the breaking of $U(1)_{\tilde{Q}}$) 
corresponds physically to a plane-wave plasma oscillation of the 
(rotated) electric charge density.  Note again that the
background condensate breaks the usual superselection rules and
allows the excitations to be coherent mixtures of states with
different hypercharges and (rotated) electric charges.

\section{Discussion}

We have shown that the homogeneous $K^0/K^+$ condensate in CFL
matter at non-zero $\mu_{\nu}$ is stable against the production
of a heterogenous, charge-separated mixed phase.  Due to the 
very short potential energy barrier between the pure $K^0$ and
$K^+$ phases, the cost of surface and Coulomb energies is too
large for charge separation to be energetically favorable.
Although we have focussed our discussion on a single point in
the $\mu$-$\mu_{\nu}$ phase diagram, the stability of the
homogeneous phase persists throughout the $C\!F\!L \, K^0/K^+$ region
identified as the ground state in Ref. \cite{kaplanreddy}.  
(As pointed out in Ref. \cite{pauloCW}, the Coleman-Weinberg
mechanism \cite{CW} may produce a small additional incentive favoring
the production of charge-separated domains.  However, this
incentive appears to be either absent or negligible for 
realistic densities relevant to neutron star cores.)

The normal modes and excitation spectra we discussed in the
previous section may be relevant for future calculations of the
neutrino scattering rates in the cooling neutron star.  Of special
importance in this context is the relatively light ($m_{\gamma}
\lesssim 10 MeV$) photon mode.  Since this excitation 
involves a spatially fluctuating density of electrons and 
charged and neutral kaons, one may expect large
cross sections for neutrino scattering.  The physics
here is similar to that discussed in Ref. \cite{nuscatter} where it
is shown that coherent neutrino scattering from droplets of kaonic
matter (whose weak charge density differs significantly from the
hadronic background) will dramatically affect neutrino scattering
rates and hence potentially observable neutrino light curves from
supernovae.  Equivalently in our problem, one expects that due to
the spatial oscillation of the electron and kaon densities,
the cross section for neutrino - photon scattering will
be large and that the copious presence of such excitations for
$T \; \gsim \; m_{\gamma} \sim 8 MeV$ may significantly increase
the time required for neutrinos to diffuse out of the proto-neutron 
star.  One might hypothesize that only the relatively early-time 
neutrino light curve would be thereby stretched out in time since,
if the dominant contribution to neutrino scattering rates indeed
comes from these longitudinal photon excitations, there will be
a relatively sudden change in the density of ``targets'' as the core
temperature drops below $m_{\gamma}$.  

Also, as discussed in Ref. \cite{vortices}, the breaking of various
$U(1)$ symmetries (hypercharge and rotated electromagnetism) by the 
mixed kaon condensate will allow for the presence of global and/or 
gauged vortices and vortons.  The present work, in determining the 
in-medium masses for the various kaonic excitations, thereby also 
fixes certain properties of kaonic vortices, {\it e.g.}, the distance 
scale over which a superconducting $K^+$ condensate at the center of 
a $K^0$ vortex will fall off with distance.  The in-medium masses of 
the various components of the (rotated) photon will also be relevant 
for a detailed description of the spatial dependence of the electric
and magnetic fields surrounding such vortices.  These vortices
may also affect the cooling history of young neutron star through
their neutrino opacities and magnetic properties.

A more detailed treatment of the various particle-like and
topological excitations above the homogeneous $K^0/K^+$ condensate,
and especially their effects on neutrino scattering and diffusion,
will be left for future work.

\begin{center}
\large{ \textbf{Acknowledgements}}
\end{center}
We thank David Kaplan, Guy Moore, and Sanjay Reddy for proposing this 
problem to us, and for many helpful discussions along the way.  The work 
is supported in part by the US Department of Energy grant
DE-FG03-00ER41132.

\newcommand{\IJMPA}[3]{{ Int.~J.~Mod.~Phys.} {\bf A#1}, (#2) #3}
\newcommand{\JPG}[3]{{ J.~Phys. G} {\bf {#1}}, (#2) #3}
\newcommand{\AP}[3]{{ Ann.~Phys. (NY)} {\bf {#1}}, (#2) #3}
\newcommand{\NPA}[3]{{ Nucl.~Phys.} {\bf A{#1}}, (#2) #3 }
\newcommand{\NPB}[3]{{ Nucl.~Phys.} {\bf B{#1}}, (#2)  #3 }
\newcommand{\PLB}[3]{{ Phys.~Lett.} {\bf {#1}B}, (#2) #3 }
\newcommand{\PRv}[3]{{ Phys.~Rev.} {\bf {#1}}, (#2) #3}
\newcommand{\PRC}[3]{{ Phys.~Rev. C} {\bf {#1}}, (#2) #3}
\newcommand{\PRD}[3]{{ Phys.~Rev. D} {\bf {#1}}, (#2) #3}
\newcommand{\PRL}[3]{{ Phys.~Rev.~Lett.} {\bf {#1}}, (#2) #3}
\newcommand{\PR}[3]{{ Phys.~Rep.} {\bf {#1}}, (#2) #3}
\newcommand{\ZPC}[3]{{ Z.~Phys. C} {\bf {#1}}, (#2) #3}
\newcommand{\ZPA}[3]{{ Z.~Phys. A} {\bf {#1}}, (#2) #3}
\newcommand{\JCP}[3]{{ J.~Comput.~Phys.} {\bf {#1}}, (#2) #3}
\newcommand{\HIP}[3]{{ Heavy Ion Physics} {\bf {#1}}, (#2) #3}
\newcommand{\RMP}[3]{{ Rev. Mod. Phys.} {\bf {#1}}, (#2) #3}
\newcommand{\APJ}[3]{{Astrophys. Jl.} {\bf {#1}}, (#2) #3}

\end{document}